\documentclass[12pt,preprint]{aastex}

\usepackage{enumerate}
\usepackage{amsthm}
\usepackage{amsmath}
\usepackage{amsfonts}
\usepackage{graphicx}

\shorttitle{Sun-quakes produced by flux rope eruption}
\shortauthors{S. Zharkov et al}
\begin{document}
\title{February 15, 2011: sun-quakes produced by flux rope eruption}
\author{S. Zharkov\altaffilmark{1}, L.M. Green\altaffilmark{1}, S.A. Matthews\altaffilmark{1}, V.V.Zharkova\altaffilmark{2}}

\altaffiltext{1}{UCL Mullard Space Science Laboratory,
          University College London, Holmbury St. Mary, Dorking, RH5 6NT  UK}
\altaffiltext{2}{{\it Department of Mathematics,} University of Bradford, Bradford, {\it BD7 1DP} UK}

\begin{abstract}
We present an analysis of the 15 February 2011 X-class solar flare, previously reported to produce the 
first sunquake in solar cycle 24  \citep{Kosovichev2011}. Using acoustic holography, we
confirm the first, and report a second, weaker, seismic source associated with
this flare. We find that the two sources are located at either end of a sigmoid
which indicates the presence of a flux rope. Contrary to
the majority of previously reported sunquakes, 
the acoustic  emission precedes the  peak of major hard X-ray (HXR) sources by several minutes.  Furthermore, the strongest hard X-ray footpoints derived from RHESSI data are found to be located away from the seismic sources in the flare ribbons. We 
account for these discrepancies within the context of a phenomenological model of a 
flux rope eruption and accompanying two-ribbon flare.
We propose that the sunquakes are triggered at the footpoints of the erupting flux rope at the start of the flare impulsive phase and eruption onset, while the main hard X-ray sources appear later at the footpoints of the flare loops formed under the rising flux rope. Possible implications of this scenario for the theoretical interpretation of the forces driving sunquakes  are discussed.

\end{abstract}

\section{Introduction}
\label{sec:intro}
Solar quakes, first observed by \citet{kz1998}, are seen as ripples
in the photosphere which move radially outwards from a source region. They are
produced as acoustic waves travel into the Sun and refract back to the
photosphere. Sunquakes are normally detected via helioseismic methods such as the construction of time-distance diagrams \citep{kz1998, Kosovichev2007, Zharkova07} or acoustic holography's egression analysis \citep{Donea1999, DL2005, ZH2011}. 
Analysis of sunquakes offers us an opportunity to explore the physical processes of energy transport in flaring atmospheres.

The theoretical prediction that sunquakes should be produced by the energy released during major solar flares \citep{wolff72} was supported by their discovery on the Sun by \citet{kz1998}. 
Following the first quake observation, further events were detected with holography techniques \citep{DL2005, Donea2006} and time-distance \citep{K2006, Kosovichev2007}. These events showed an association either with X-class flares, such as those on 28 and 29 October 2003 \citep{DL2005}, or with M-class flares such as that on 9 September 2001 \citep{Donea2006}. However, not all flares show seismic activity as concluded by \citet{Besliu2005} and \citet{Donea2006list}, who reported a catalogue of only 17 flares of X and M class with measurable seismic activity detected by either holographic or time-distance approaches.
These observations posed a question about how the energy and momentum are transported to the solar surface and interior, in order to produce sunquakes and why some of the most powerful flares often do not  deliver seismic signatures. 

Many of the previously detected seismic ripples and acoustic sources associated with flares were found to be co-spatial with the hard X-ray (HXR) source locations  \citep[the vast majority reported in][]{Besliu2005, Donea2006list}, while in some flares the seismic sources were co-spatial with $\gamma$ sources \citep{Zharkova07, Kosovichev2007}. These cases support the idea of sunquakes being produced by hydrodynamic shocks induced by the ambient plasma heating either by electron \citep{kz1998, Kosovichev2007} or proton beams \citep{Zharkova07}. \citet{Zharkova07} suggested that apart from high energy proton or electron beams, jet protons with quasi-thermal energy distributions can also be the source of acoustic emission and seismic ripples. Such jets, with maxima shifted to several MeV, can be ejected from a current sheet during the magnetic reconnection process  \citep[see for example][and references therein]{Barta2011}.

\citet{Donea2006} noted that in many cases considered by \citet{Besliu2005} the
location of acoustic sources were co-spatial with, and had an energy range similar to the white light emission from
these flares. As a result, it was proposed that back-warming heating of the photosphere by
the overlying radiation from the corona and chromosphere was the source of
acoustic emission \citep[see for example][and references therein]{Donea2006, donea11}. 

On the other hand,  \citet{Hudson08} noted that the energy associated with the reconfiguration of the magnetic field  
during a flare,  the so called McClymont jerk, can easily account for the energy required for a sunquake.  
Observationally this reconfiguration is seen 
in the line-of-sight magnetic field in the photospheric as an abrupt and permanent
magnetic field change \citep{kz01,Zharkova2005, Sudol2005}. The energy released during an irreversible magnetic field change was found to be sufficient to account for the whole flare emission  \citep{Zharkova2005}.  \citet{Hudson08} suggested that the seismic emission is initiated directly by these magnetic pulses in a form of magneto-acoustic waves \citep{cally00, Martinez2008b, Martinez2009mnras}. However, in a study of two flares with sunquakes \cite{Martinez2009mnras} found inconclusive results.

The February 15, 2011, X2.2 class flare was the first in the much delayed rising activity phase of the new solar cycle 24.  \citet{Kosovichev2011} has reported that the flare produced a sunquake clearly seen as propagating circular ripples in running difference filtered velocity images of the surface.  We present a new study that detects more seismic sources in this flare using holographic methods, and an investigation of their dynamics by taking into account the morphology of the active region and the occurrence of a coronal mass ejection in association with the flare.  We describe the data in section \ref{obser}, report on the photospheric and coronal observations in section  \ref{sec:results} and discuss possible sunquake production in the context of a flux rope eruption and two-ribbon flare model in section \ref{sec:conclusions}. 

\section{Observations and data processing} \label{obser}

\label{sec:observations}
We use full disk SDO/HMI observations, from which the helioseismic datacubes are extracted by remapping and de-rotating the region of interest using Postel projection and Snodgrass differential rotation rate. In this way we obtain 45 second cadence datacubes of HMI line-of-sight velocity, magnetogram and intensity images. 
The spatial resolution for the remapped data is $0.04$ heliographic degrees per pixel.
The centre of the extracted region is located at $20^\circ$ latitude South and $11.75^\circ$ longitude to the East. 
The series starts at  00:59 UT 15 February 2011 and runs for three hours. 

From the velocity running difference datacubes we measure the acoustic egression following the processing as outlined in \citet{Donea1999, LB2000, donea2000}. The Green's functions are obtained by solving the non-magnetic wave equation for monochromatic point source via geometric optics.
To study the flare onset
we use the data from SDO's Atmospheric Imaging Assembly (AIA) instrument obtained for the same period at $1700$ \AA\ and $94$ \AA\ wavelengths.

The hard X-ray data presented in this paper are derived from RHESSI \citep{lin02}, with Hinode XRT \citep{Golub07} providing data for soft X-ray information. RHESSI observed the flare from the pre-cursor phase beginning at ~01:27 UT until
02:30 UT, covering the entire impulsive phase. We used the CLEAN algorithm to
produce images at between 20 and 40 second cadence covering the duration of the flare. 

 \section{Results} \label{sec:results}
\subsection{Active region features} \label{features}
The sunquake occurred in NOAA active region 11158 which began emerging in the Eastern hemisphere on the 10 February 2011. Two bipoles emerged side-by-side creating a complex multipolar region.  As the active region evolved through both emergence and cancellation events the coronal loops became increasingly sheared, and by late 14 February the loops in the northern part of the active region showed a forward S shaped sigmoidal structure in soft X-rays and EUV emission  (Figures \ref{fig:eg_snapshots}-\ref{fig:eg_rhessi}). The occurrence of a sigmoid in the active region gives strong support to the presence of a magnetic flux rope at this location \citep{GreenKliem2009}. The sigmoid formed along a polarity inversion line, where the flux cancellation, provides the mechanism by which the helical field lines of the flux rope can be formed from a sheared arcade \citep{vanBall1989, GreenKliem2011}.

On 15 February 2011, AR 11158 produced a coronal mass ejection (CME) with the
eruption being evidenced by the rise of a linear loop-like feature in AIA 94 $\AA$ data (see Figure \ref{fig:changes} and online movie). The CACTus CME catalogue observed a halo CME in LASCO C2\footnote{the CME is actually listed as three separate CMEs: numbers 34, 35 and 36 at {\rm http://sidc.oma.be/cactus/catalog/LASCO/2\_5\_0/qkl/2011/02/}}.  The halo CME was first seen in LASCO C2 above the occulting disc to the south-west at 02:24 UT and had a plane-of-sky velocity of between 274 and 469 km/s. The true CME velocity may have been considerably higher as the CME originated near Sun centre.
 
The CME was accompanied by an X2.2-class two ribbon flare and an
EUV wave. The flare impulsive phase as seen in the GOES 1.0 to 0.8 \AA\ soft X-ray data occurs between 01:46 and 01:56 UT. Integrated HXR emission was observed with
RHESSI up to approximately 100 keV.
$\gamma$-ray line emission was not observed by RHESSI during the flare.  RHESSI images obtained with CLEAN procedure were used
to provide spatially resolved light-curves of the HXR sources in the vicinity of the
egression sources and the main flare ribbons (Figure \ref{fig:eg_rhessi}).

\subsection{Two seismic sources} \label{seismic}
\begin{figure*}
\centering
\includegraphics[width=17.5cm]{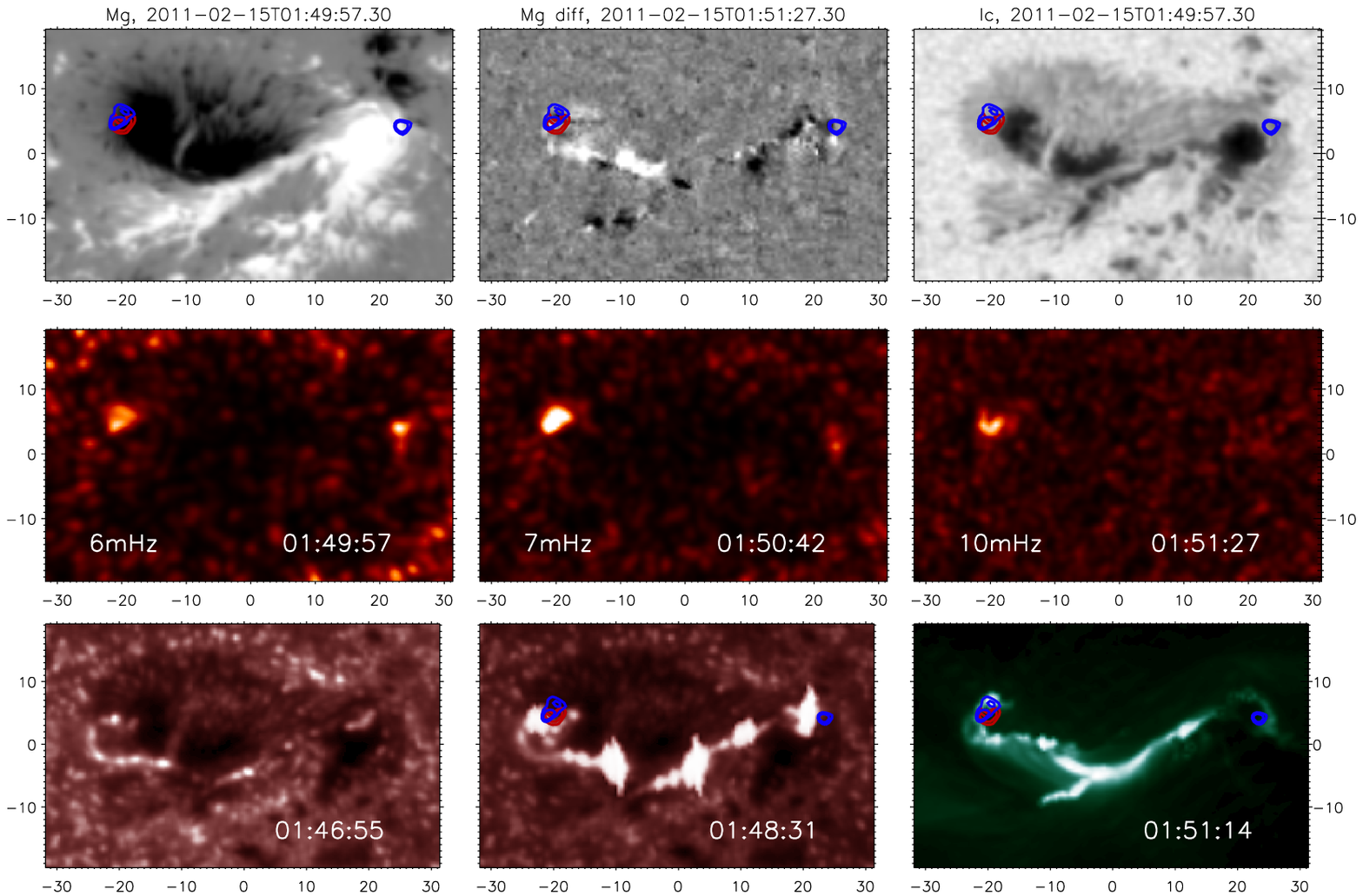} 
\caption{First row: magnetogram, magnetogram difference and intensity images .
Second row: egression power snapshots at different frequencies taken on 2011-02-15.
Third row (left to right): two AIA 1700 \AA~snapshots and an AIA 94 \AA~image showing flare ribbons.
On all images, the blue contours are 2011-02-15 01:49:57 6mHz egression power
snapshot at 2.5 and 3 times quiet Sun egression power. Red contours are the 10mHz egression power (same time) at 3 and 4 times quiet Sun egression power.
The images are remapped onto heliographic grid, the distance is plotted in Megameters.
\label{fig:eg_snapshots}}
\end{figure*}

Computed egression power snapshots for 6, 7 and 10 mHz frequency bands taken around the times of the peak in the acoustic emission are shown in Figure \ref{fig:eg_snapshots}. The data are scaled by the mean quiet sun egression power value at each frequency and saturated at factor 5 for better contrast. At 6 mHz one
can clearly see the two strong acoustic sources located in the Eastern and Western  parts of the image. The locations of the sources are plotted as contours over magnetogram and intensity images in the top of Figure \ref{fig:eg_snapshots}. The Eastern source (Source 1), which corresponds to the acoustic source reported by \citet{Kosovichev2011}, is larger and stronger, clearly seen in all frequency bands of computed egression snapshots. The Western source (Source 2) is considerably smaller, and best seen in 6 mHz band becoming faint at 8 mHz. 

In order to check the significance of acoustic sources, following  \citet{Donea1999,  donea2000, MZZ2011} we have performed rms analysis by spatially integrating egression power over a $ 140 {\rm \  Mm}^2$ region obtained via morphological dilation of the egression kernels. The 6 mHz results are shown in the bottom row of Figure \ref{fig:variations}. It is clear that the acoustic signal at Source 1 is very strong, exceeding the mean value of the series by a factor of up to 2.9. While Source 2 is clearly weaker, the signal at 6 mHz band exceeds the mean by a factor of 2.4. 

\begin{figure*}
\centering
\includegraphics[width=17.5cm]{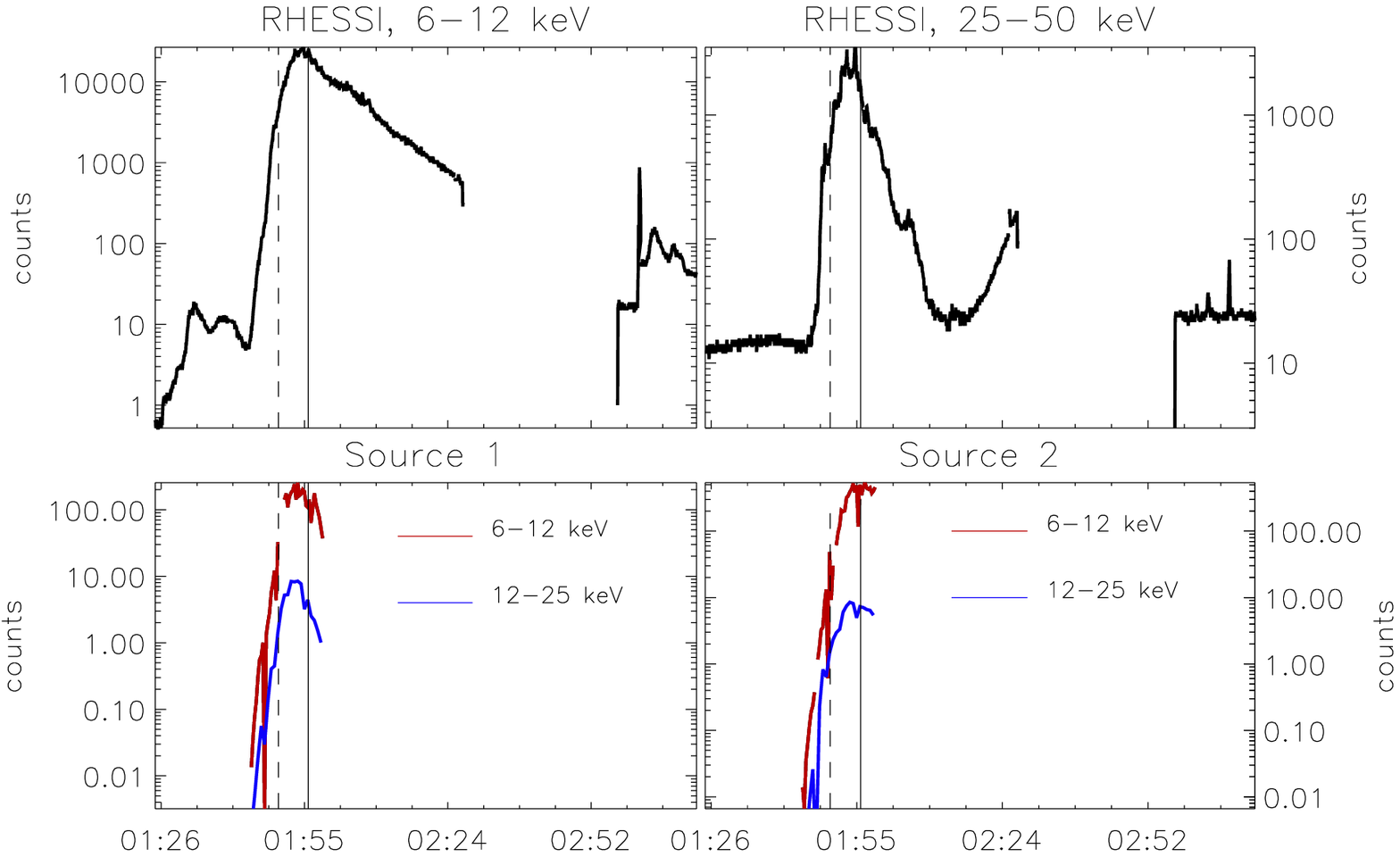} 
\includegraphics[width=17.5cm]{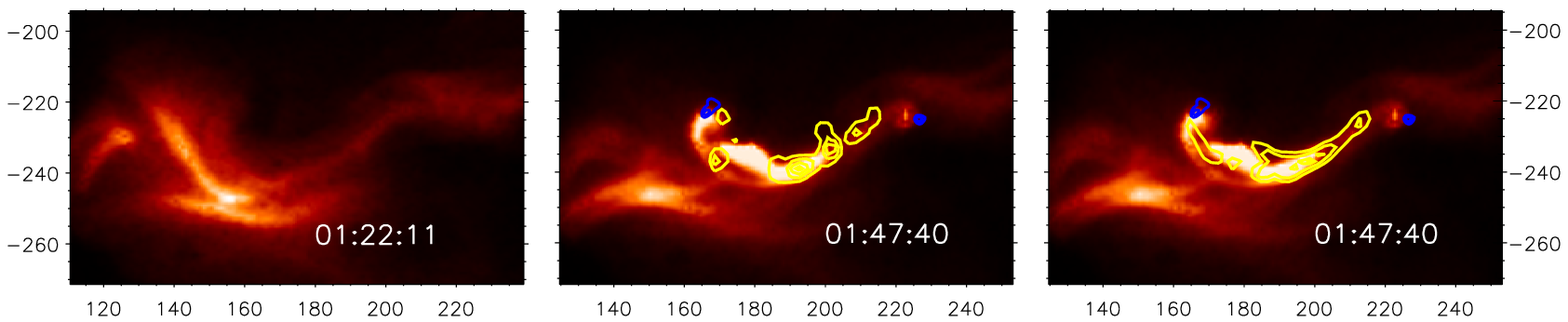} 
\caption{
The top row are RHESSI counts (over the whole region). The next row are RHESSI data integrated over egression sources. The vertical lines correspond to 01:50 UT and 01:56 UT.
Bottom row (left to right): Hinode XRT image showing the sigmoid and
RHESSI contours for the following energy ranges: 12-25 keV(middle plot), 6-12 keV(right).
The arcsecond coordinates are plotted along $x$- and $y$-axis. The red and blue contours are as in Figure \ref{fig:eg_snapshots}.
\label{fig:eg_rhessi}}
\end{figure*}

The largest magnetic, intensity and velocity variations associated with the
CME/flare occurred along the flare-loop footpoints away from the seismic sources.
The strongest HXR emission is also situated primarily at the site of the flare loops (Figure \ref{fig:eg_rhessi}), with the peaks apparently corresponding to the sites of maximum magnetic field changes, some distance away from the seismic sources. Weaker HXR emission is seen at the sites of the seismic sources, primarily in the 6-12 and 12-25 keV energy ranges.

At the same time, the evolution of velocity, intensity and magnetic field, presented in Figure \ref{fig:variations}, where the values are integrated over the acoustic kernels derived from the 6mHz egression snapshot, show significant changes associated with the flare. In the velocity data the downward propagating shocks are seen around 01:49-01:50 UT at both sunquake locations. This is accompanied by a small increase in the intensity and the start of a gradual process of the line-of-sight magnetic field change that is permanent and takes place over several minutes until the time of the peak HXR emission, 01:53-01:54 UT (see Figure \ref{fig:eg_rhessi}, top row). However, the presence of the weak HXR emission indicates that such transients could occur in these locations, perhaps in a single pixel \citep{Martinez2009mnras}, thus we cannot  completely rule them out.
 
\begin{figure*}
\centering
\includegraphics[width=7.9cm]{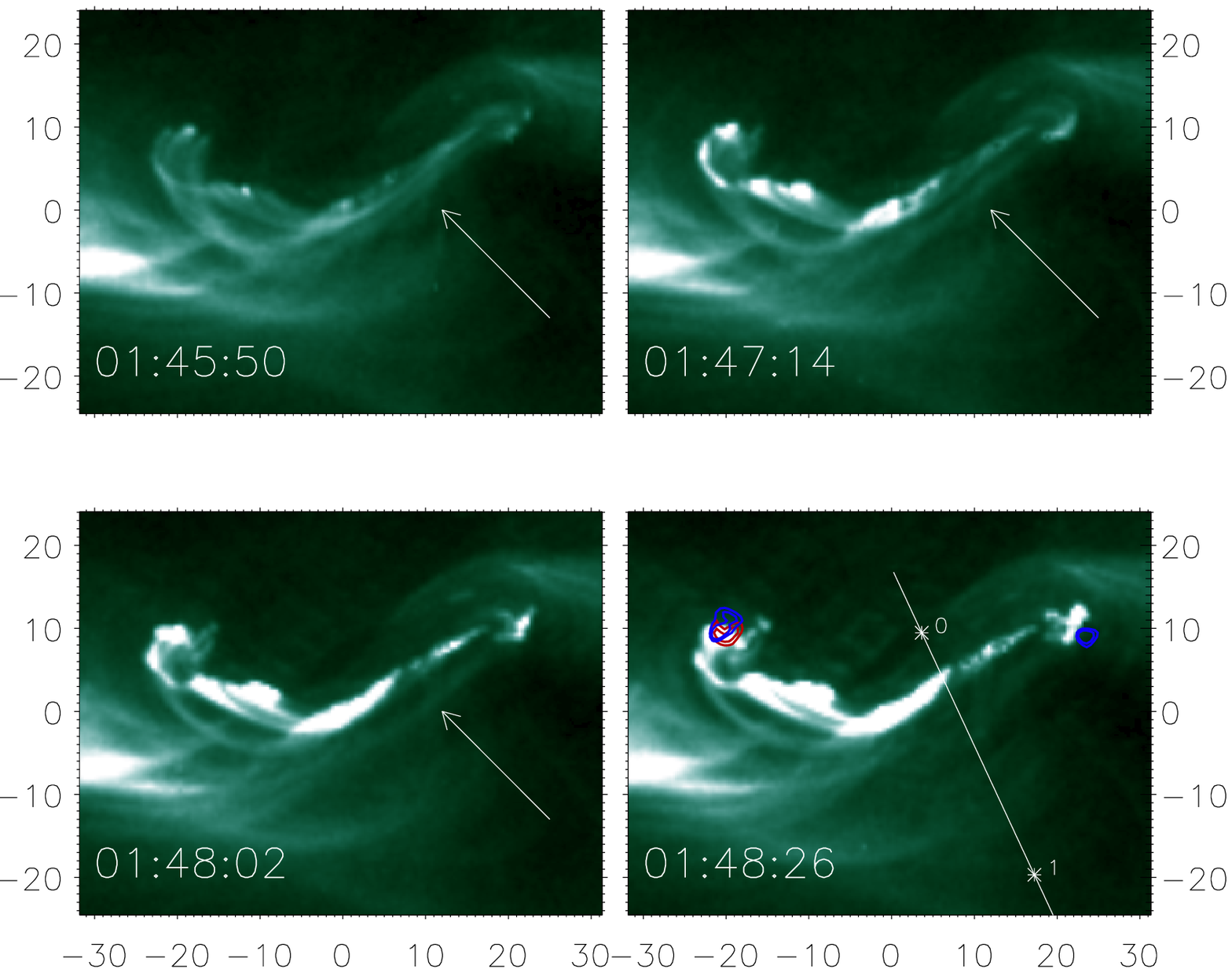} 
\includegraphics[width=7.9cm]{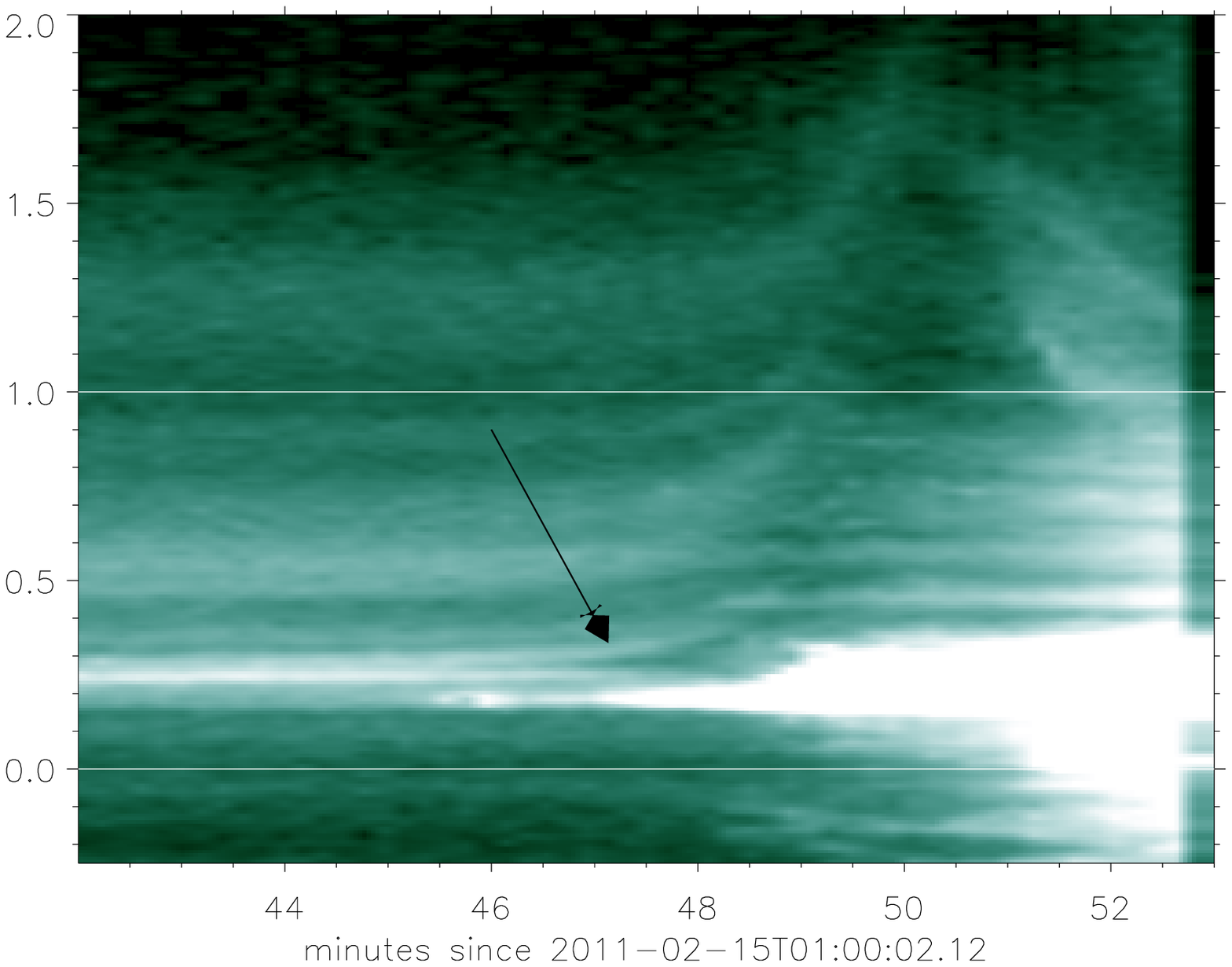} 
\caption{AIA 94 \AA~data showing the evolution of the coronal structure at the onset of the flare, CME and sunquake. A 'loop-like' feature is
seen to erupt away from the body of the sigmoid {\em(white arrow in left-hand panels)} . The stack-plot
shows a slice across the sigmoid in the direction of the motion of erupting structure. The erupting structure is indicated by the black arrow in
the stack-plot {\em (right)} obtained along the line shown in the image at 01:48:26 UT.  Distance along the line is plotted along the $y$-axis in the stack-plot, with values for 0 and 1 indicated in the snapshot.
\label{fig:changes}}
\end{figure*}

\begin{figure*}
\centering
\includegraphics[width=16.cm]{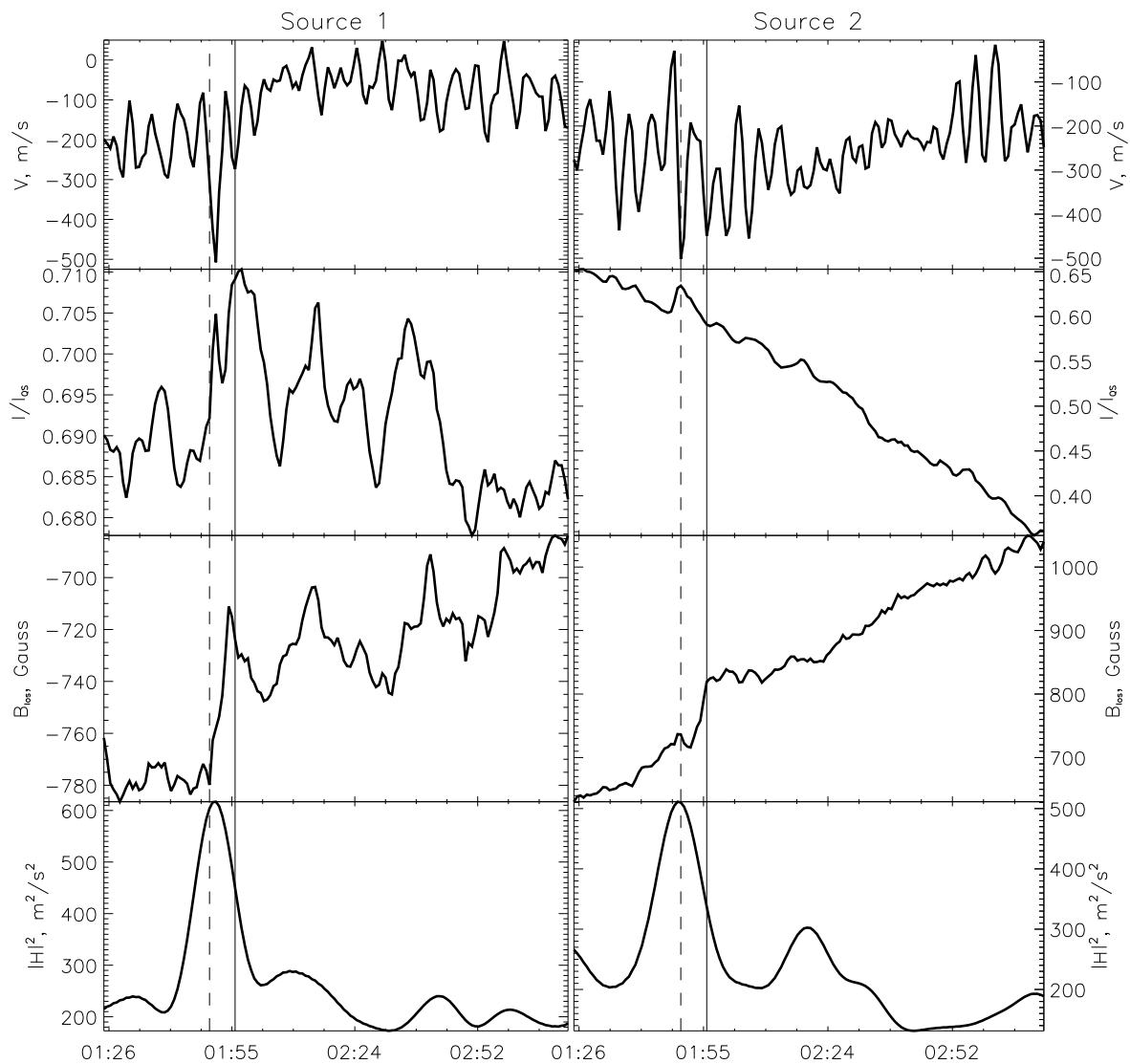} 
\caption{
Velocity, intensity and magnetic field variations integrated over 6 mHz egression kernels (using 2.5 factor of quiet Sun egression value as a threshold). The bottom plots are egression rms at 6mHz.  The vertical lines correspond to 01:50 UT and 01:56 UT.
\label{fig:variations}}
\end{figure*}

 
 In spite of the frequency filtering of the acoustic egression power that limits the measurements of exact timing of the seismic emission  \citep{Donea1999},  the peak emission at both the sources is clearly seen between 01:48-01:50 UT, which is in agreement with the time-distance ridge results published in \citet{Kosovichev2011} and velocity transients seen at these locations in HMI data.  A comparison with the RHESSI lightcurve data (Figure \ref{fig:eg_rhessi}, top row) reveals that the quakes occur during the early stage of the impulsive phase of the flare, and before the peak in HXR. The intensities of HXR emission in the locations of ribbons exceeds 100 times  those in the locations of seismic so¤urces (compare the two upper rows in Figure \ref{fig:eg_rhessi}). Furthermore, Figures \ref{fig:eg_rhessi} (bottom row) and \ref{fig:changes} (01:48:26UT snapshot) show that the two seismic sources are co-spatial with the curved ends of the sigmoid seen in soft X-ray and EUV emission. These are also the locations where EUV changes are first seen around 01:46 UT during the onset of the flare as shown in Figures \ref{fig:eg_snapshots} and \ref{fig:changes}. 


Thus, the observations indicate that the seismic sources occur close to the edge of the flux rope endpoints, and in the region and where the energy release during the CME is expected to take place \citep{titov99}.  The onset of the CME is seen in the lower corona as the rise of a loop-like structure shown in Figure \ref{fig:changes}. The Western end of the loop-like structure appears to be rooted in, or very close to, the Western seismic source. The loop-like feature first undergoes a slow rise phase followed by a rapid acceleration around 15 February 01:48 UT (see bottom panel and arrow in Figure \ref{fig:changes}).

\section{Discussion and conclusions} \label{sec:conclusions}

Unlike most previously detected sunquakes, the two seismic sources detected for
this event are located away from the sources of very bright HXR emission. This
indicates that the sunquakes are not forming at the location of the strongest
particle precipitation. Furthermore, both sunquakes apparently occur early on
in the impulsive phase of the flare, about three to five minutes before the
hard X-ray emission reaches its peak in all of the energy bands. 
The key to
understanding the quakes produced in this event seems to be the magnetic structure of the sigmoid.

The observation of the sigmoid and the erupting loop-like feature gives strong support for the presence of a flux rope before and during the eruption \citep{mckenzie08,aulanier10,GreenKliem2011}. We propose that initial small-scale energy release takes place in a quasi-separatrix layer (QSL) formed at the interface between the flux rope and the surrounding arcade field. This energy release produces the bright patches that later increase in intensity and area to form the flare ribbons. Their J -shapes represent the location of the intersection of this QSL with the lower solar atmosphere \citep{demoulin96}. 

\begin{figure*}
\centering
\includegraphics[width=16.2cm]{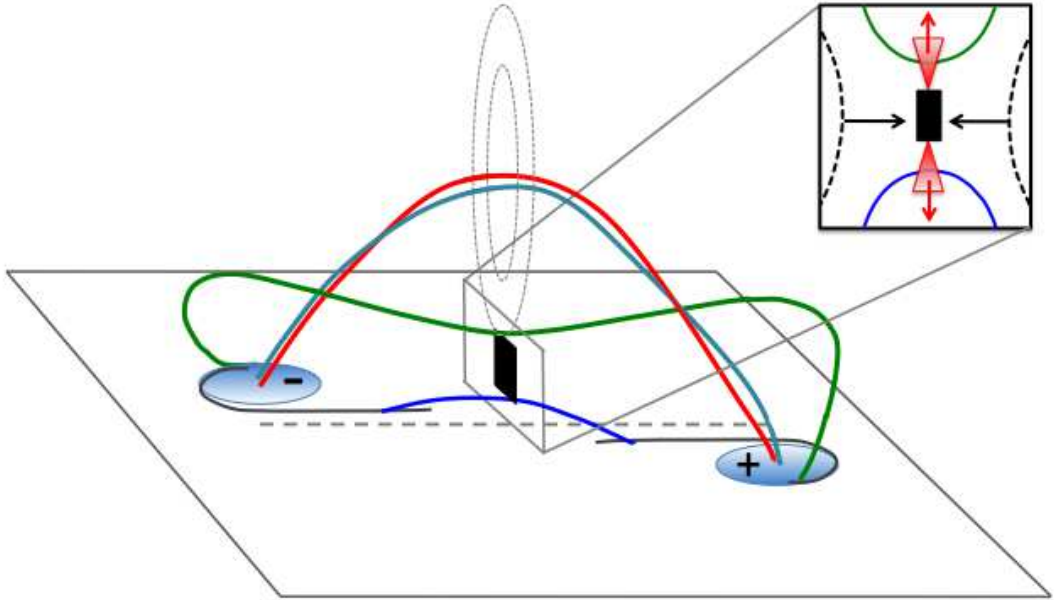}
\caption{The flux rope scenario: red line represents a field line at the axis
  of the erupting flux rope and the teal line a field line close to the axis
  with a right-handed twist. The flux rope cross section is illustrated by the
  two grey, dashed ovals. As the flux rope erupts, overlying sheared field
  lines reconnect in the current sheet formed under the rope producing two sets
  of field lines; short flare arcade field lines (shown in blue) and longer
  field lines that become part of the flux rope body, making roughly one turn
  about the flux rope axis (shown in green). Associated to this reconnection are particle jets represented in red in the inset box. At the photosphere the polarity inversion line (grey dashed) and flare ribbons (grey solid) are indicated. 
\label{fig:model}}
\end{figure*}

As the magnetic configuration evolves, the flux rope reaches a stage where it is no longer in equilibrium and a rapid acceleration phase sets in. This is schematically presented in Figure \ref{fig:model}. At this point, the X-line under the flux rope collapses into a current sheet facilitating reconnection under the erupting structure and leading to the two-ribbon flare seen in SDO and Hinode data.
This reconnection brings in surrounding arcade field and produces the flare loops seen in SXR and EUV emission (shown in blue in Figure \ref{fig:model}), and also forms helical field lines that wrap around the flux rope body \citep{titov99} and further enhance the speed of the eruption \citep{zhang04} (shown in green in Figure \ref{fig:model}, see also Figure 1 in \citep{Shibata95}). To confirm this scenario, we note that the footpoints of the newly formed field lines which wrap around the body of the flux rope merge with the flaring ribbons during the reconnection process as shown by \citet[][their Figure 4]{demoulin96}. Therefore, the J-shaped ribbons represent the end points of all field lines involved in the reconnection. 

The timing and the location of the egression emission sources indicate that not only are the seismic  responses generated at the feet of the erupting flux rope,  but that they are generated close in time to the flare/CME onset. 
The absence of
strong HXR and white light emission at the seismic sources appears to rule out
back-warming as the physical mechanism of quake excitation.  Absence of strong HXR is in line with \citet{Shibata95} model for two ribbon flare.
The question then arises as to what  mechanism(s) generate each of the seismic sources given their similar timing but different powers, and why they do not appear at the footpoints of the flare arcade. In order to explore these points we consider two avenues: hydrodynamic shocks and magnetic re-structuring.

Firstly, consider the current sheet under the erupting flux rope where two types of particles are ejected:  quasi-thermal particles of separatrix jets ejected from the current sheet sides by magnetic diffusion, and electrons and protons dragged from the corona and accelerated in the current sheet to sub-relativistic energies  \citep{ZharkovaAgapitov, Siversky2009JPlPh}. These high energy particles can be ejected either separately or as mixed beams down the field lines of the flare arcade and along the field lines that wrap around the flux rope. Because the current sheet associated with this flux rope model is located in the corona, the density of  the dragged-in and accelerated particles is relatively low. Thus, even if particles are accelerated to high energies, they will not produce noticeable HXR or $\gamma$-ray emission. The jet particles combined with high energy electrons and protons could produce a mild chromospheric evaporation into the corona resulting in UV emission and a strong hydrodynamic shock in the lower chromosphere/photosphere. These shocks, in turn, could produce seismic emission as shown by hydrodynamic simulations by \citet{Zharkova07}. 
 
Secondly, we observe a clear and abrupt permanent change in the magnetic field in the strongest seismic source. An abrupt change is also seen in the weaker source, although the field in this region shows a continual increase that begins before the CME/flare onset, making the interpretation of the abrupt change more complex. We also note that both egression sources are located in the penumbral field, which is relatively inclined toward the horizontal before the eruption and must become more vertical when the flux rope erupts and expands. These observations are consistent with the the idea of a magnetic jerk in response to coronal restructuring produced by the CME, but more investigation is required in order to fully understand how these changes relate to the observed seismic emission. In conclusion, the observation of two seismic sources located at the ends of the erupting flux rope highlights the importance of understanding the role of magnetic field topology in the generation of seismic emission, a factor not currently included in existing models.  Further studies with high resolution and high cadence data are now needed to determine the mechanisms behind sun-quake production in the context of an erupting magnetic configuration and associated particle transport, rather than only focussing on flare related mechanisms. 

\acknowledgements
The authors thank Dr B. Kliem for many useful discussions.

\end{document}